# Title

Symmetry violation-driven hysteresis loops as measurands for noise-resilient sensors

# Authors

Arunn Suntharalingam[1], Lucas Fernández-Alcázar[1,2,3], Pablo Fabián Wagner-Boián[3], Mattis Reisner[1], Ulrich Kuhl[1,4], Tsampikos Kottos[1, *]

[1]Wave Transport in Complex Systems Lab, Department of Physics, Wesleyan University, Middletown, CT 06459, United States

[2]Institute for Modeling and Innovative Technology, IMIT (CONICET - UNNE), W3404AAS, Corrientes, Argentina

[3]Physics Department, Natural and Exact Science Faculty, Northeastern University of Argentina, W3404AAS, Corrientes, Argentina

[4]Université Côte d'Azur, CNRS, Institut de Physique de Nice (INPHYNI), 06200, Nice, France

[*]Corresponding Author: Tsampikos Kottos (tkottos@wesleyan.edu)

# Abstract

Sublinear resonant deviations from an exceptional point degeneracy (EPD) has been recently promoted as a sensing scheme. However, there is still an ongoing debate whether the sensitivity advantage is negated by an increase in fundamental noise – especially when active elements induce self-oscillations. In this case, nonlinearities are crucial in stabilizing amplifying modes and mitigating noise effects. A drawback is the formation of hysteresis loops that signal a transition to unstable modes. This can only be alleviated by precise cavity symmetry management. Here, utilizing two coupled nonlinear $RLC$ tanks with balanced amplification and attenuation, we demonstrate that an explicit symmetry violation, induced by sweeping the resonant detuning of the $RLC$ tanks, reveals a hysteresis loop near the EPD whose width scales sublinearly with the inter-tank coupling. Our proposal re-envisions this disadvantageous feature as a sensing protocol with diverging sensitivity, enhanced signal-to-noise ratio, and self-calibration without requiring delicate symmetry control. As such, it opens new avenues in metrology as well as for optical/RF switching and triggering.

## I. Introduction

The eigenfrequency Riemann surfaces (RS) of non-Hermitian Hamiltonians has unveiled a wealth of counter-intuitive phenomena with various technological ramifications. Exceptional point degeneracies (EPDs) [1] have undoubtedly been the tip of the spear for many of these new developments [2]-[5]. These are resonant singularities occurring in the parameter space of non-Hermitian cavities, where $M \geq 2$ eigenfrequencies and the corresponding eigenvectors coalesce. In their proximity, the RS self-intersect, enforcing a fractional perturbation expansion of the eigenmodes $\Delta\omega \equiv \omega - \omega_{EPD} = \sum_n c_n \delta^{n/M}$, where $\delta$ is a small system-parameter variation and $\omega_{EPD}$ marks the EPD frequency at which the RS splits. This sublinear frequency response has been recently proposed for enhancing the detection sensitivity [6]-[15], generating nonreciprocity [16]-[21], chiral emission [22]-[23][25], and more.

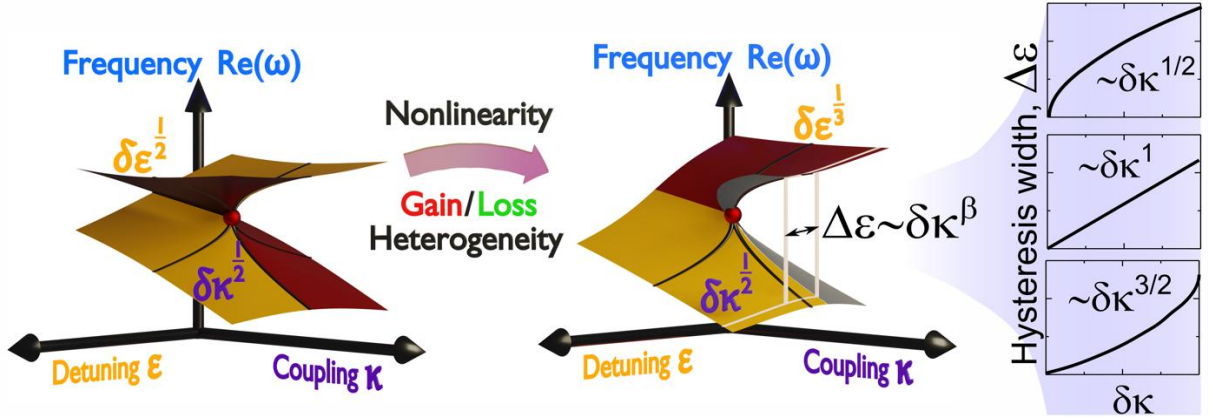

*Figure 1: **Nonlinearity-induced hysteresis loop scaling via a metamorphosis of the Riemann surfaces.** The Riemann surfaces (RSs), of a linear (left) and a nonlinear (center) system of two coupled oscillators with heterogeneous gain and loss distribution, in the detuning ε and the coupling κ parameter space. Nonlinearities violate the isotropy of the RS leading to different fractional ε and κ expansions away from the EPD (red sphere). Importantly, a stable (experimentally accessible) RS can undergo a "metamorphosis" to an unstable (experimentally inaccessible) RS indicated as a grey semitransparent surface (center). Such transformations give rise to the formation of hysteresis loops, where two stable supermodes coexist for the same parameter values. The width of the hysteresis loop Δε is marked by the parameters where the unstable-to-stable transition takes place and its scaling with the orthogonal Riemannian parameter δκ is controlled by the specifics of the nonlinear interactions (right).*

Some of these studies include amplifying elements which may trigger nonlinear mechanisms, e.g. lasing [26]-[29]. Applications that transpire from the synergy of nonlinearities and non-Hermitian degeneracies are robust wireless power transfer [30], asymmetric transport [31], engineered bistability for noise-aided state switching [32], nonlinear control of non-Hermitian topological states [33] and hypersensitive sensing based on lasing action [11][12]**Error! Reference source not found.**-[35]. At the core of the above achievements is the concept of nonlinear EPDs (NLEPDs) whose order $M$ is determined by the number of nonlinear supermodes (the fixed points of the dynamical equations) that simultaneously coalesce [35]-[41]. Curiously, the order of degeneracy of the NLEPDs can be larger than the degrees of freedom of the system [37]-[39]. The properties of NLEPDs arise from the underlying dynamical symmetries and their stability characteristics. As opposed to the linear EPDs, where the eigenbasis collapse leads to signal-to-noise ratio (SNR) suppression, the NLEPDs are characterized by an opposite tendency – thus constituting them promising candidates for sensing purposes.

While (saturable) nonlinearities are crucial in stabilizing amplifying modes and mitigating noise effects, their presence comes with several drawbacks. The most prominent one is the formation of hysteresis loops in the parameter space that signal the emergence of unstable nonlinear supermodes from their stable counterparts as a system parameter changes. A consequence of this "metamorphosis" is the transformation of Riemannian sheets from stable (that can be experimentally probed) to unstable ones (that are directly inaccessible to experiments), see grey RS in Fig. 1. When this phenomenon occurs in the proximity of NLEPDs, we may lose the sensitivity advantage. This deficiency can only be alleviated by precise cavity-symmetry management which enforces the continuation of the stable nonlinear supermodes (stable fixed points) all the way up to the NLEPD as the perturbed parameter varies (see black lines in Fig. 1). Obviously, fine-tuning of control parameters is always a challenge – particularly under harsh operational conditions.

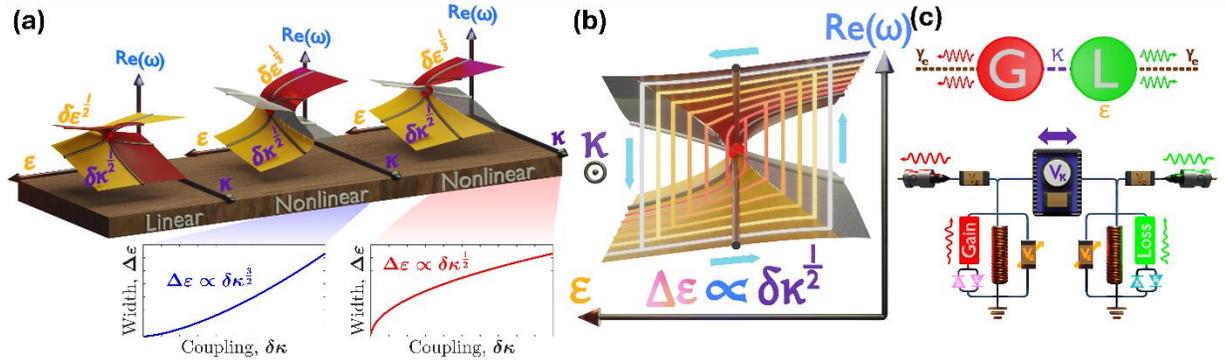

*Figure 2. Shaping the nonlinear Riemann surface for sensing applications. (a) Top: Left – the circuit configuration has $(\chi_1, \chi_2) = (0,0)$; Middle – a nonlinear configuration with $(\chi_1, \chi_2) = (1,0)$; Right – another nonlinear configuration with $(\chi_1, \chi_2) = (1,1)$. Different combinations of the nonlinear coefficients, $\chi_{1(2)}$, can modify the shape and stability of the supermode frequencies ω that form the Riemann surface. The real part of the stable (unstable) nonlinear eigenfrequencies $\Re(\omega)$ are indicated with the solid red and yellow (transparent grey) surfaces as a function of the coupling κ and resonant frequency detuning ε. The position of the exceptional point degeneracy (EPD) on each of the three surfaces are indicated by a red sphere. The solid black (grey) lines indicate the parametric evolution of the nonlinear eigenfrequencies that crosses the EPD point for fixed $\varepsilon = \varepsilon_{EPD}$ ($\kappa = \kappa_{EPD}$) when κ (ε) varies. Bottom: The different scaling of the hysteresis width, Δε, with respect to δκ for the two nonlinear Riemann surfaces corresponding to the configurations $(\chi_1, \chi_2) = (1,0)$ and $(\chi_1, \chi_2) = (1,1)$, behaves as $\Delta\varepsilon \propto \delta\kappa^\beta$ with $\beta = 3/2$ and $1/2$, respectively. (b) A zoomed face-on view of the Riemann surface with $(\chi_1, \chi_2) = (1,1)$. The colored solid lines are hysteresis loops (see light blue arrows for ε - directional sweeps) which intersect the respective Riemann surface at different coupling values, δκ, demonstrate the presence of a bistable domain. The width of the hysteresis loop, Δε, grows sub-linearly away from the EP as $\Delta\varepsilon \propto \delta\kappa^\beta$ with $\beta \approx 0.5$. (c) Top: Simplified schematic of the active dimer, with gain, loss, and nonlinearities that is modelled. Bottom: Schematic of the experimental implementation of the theoretical analogue.*

Here, instead, we exploit appropriately engineered hysteresis phenomena in the proximity of NLEPDs to develop a sensing protocol with diverging sensitivity, enhanced signal-to-noise ratio, and self-calibration without requiring delicate symmetry control – hence turning this "disadvantageous" feature into the main operational principle of a new family of sublinear hypersensitive sensors. We have implemented this sensing proposal using an experimental platform consisting of two coupled $RLC$ tanks with self-induced balanced amplification and attenuation enforced by appropriately chosen (saturable) nonlinearities. An explicit parity-time symmetry violation induced by sweeping the $RLC$ resonant detuning (between the tanks) reveals a hysteresis loop near the NLEPD whose width scales as square-root with the inter-tank coupling – allowing us to utilize it as a novel sublinear sensing measurand. It is important to stress that the (sublinear) power exponent is controlled by the nonlinear interactions that mold the structure of the linear RS near the NLEPDs – thus, making the nonlinearities an important design element (see Fig. 2a). As opposed to other NLEPD protocols [36][37][38][39], the hysteresis-width scaling scheme is intrinsically self-referenced, i.e., there is no need for an external reference to suppress or eliminate thermal drifts, etc. Specifically, the former protocols require a continuous evaluation of the critical parameter values for which NLEPD occurs, due to thermal drifts and other noise effects. Instead, our scheme is agnostic to the parameter values of NLEPD since it only requires information about the hysteresis width which can be acquired by sweeping an independent parameter (resonant detuning between the two tanks) at each variation of the inter-tank coupling (measurand). Furthermore, our scheme can be implemented using various observables e.g. the nonlinear eigenfrequencies or the self-oscillation amplitudes/fields of each $RLC$ resonator. Finally, although our experimental platform is on the order of centimetres (for adjustability), its design is easily implementable onto smaller and cost-effective integrated sensors. Our proposed scheme opens opportunities for applications beyond sensors, such as optical and RF switches and Schmitt triggers.

## II. Theoretical Modeling

First, we demonstrate the structural changes of the RS due to the presence of nonlinearities. To this end, we use a temporal coupled mode theory (TCMT) modeling [36], describing two nonlinear modes/resonators – one involving an amplification mechanism (gain) and another one incorporating an attenuation mechanism (loss). Such a model can be implemented in a variety of physical frameworks, e.g., coupled optical microcavities, coupled acoustic cavities, and radio frequency (RF) circuits. The dynamics of the modal field amplitudes $\vec{a} = (a_1, a_2)^T$ (normalized such that $|a_{1(2)}|^2$ is the modal energy) is given by

$$i\frac{d}{dt}\begin{pmatrix}a_1\\a_2\end{pmatrix} = \widehat{H}\begin{pmatrix}a_1\\a_2\end{pmatrix} = \begin{pmatrix}\omega_1 + i\gamma_1(|a_1|^2) & \frac{\kappa}{2}\\ \frac{\kappa}{2} & \omega_2 - i\gamma_2(|a_2|^2)\end{pmatrix}\begin{pmatrix}a_1\\a_2\end{pmatrix} \quad (1)$$

where $\widehat{H}$ is the effective Hamiltonian that describes our system, $\omega_{1(2)}$ is the natural frequency of the gain(loss) mode, $\kappa$ is the coupling strength between the two modes/resonators, and $\gamma_n = \gamma_n(a_n) = \gamma_1^{(0)} + (-1)^n \chi_n |a_n|^2$ is a (local) nonlinearity which depends on the field intensity in the $n = 1,2$ resonator. The functional form of $\gamma_n$, has been chosen such that $\chi_1 \neq 0$ guarantees that the nonlinear gain $\gamma_1$ saturates, thus, avoiding unbounded solutions where the fields diverge $a_n \to \infty$. It is convenient to consider the polar representation of Eq. (1) obtained through the coordinates transformation $a_n(t) = A_n(t)e^{i\varphi_n(t)}e^{-i\omega t}$ where $A_n > 0$ are the magnitudes and $\varphi_n$ the phases of the field's amplitude in resonator $n = 1,2$ (with $A_n, \varphi_n \in \mathbb{R}$). Their evolution follows the following set of coupled differential equations

$$\begin{cases} \dot{A}_1 = \gamma_1(A_1) \cdot A_1 + \left(\frac{\kappa}{2}\right) A_2 \sin(\varphi) \\ \dot{A}_2 = -\gamma_2(A_2) \cdot A_2 - \left(\frac{\kappa}{2}\right) A_1 \sin(\varphi) \\ \dot{\varphi}_1 = \omega - \omega_0 - \frac{\varepsilon}{2} - \left(\frac{\kappa}{2}\right) \frac{A_2}{A_1} \cos(\varphi) \\ \dot{\varphi}_2 = \omega - \omega_0 + \frac{\varepsilon}{2} - \left(\frac{\kappa}{2}\right) \frac{A_1}{A_2} \cos(\varphi) \end{cases} \quad (2)$$

where $\omega_0 \equiv \frac{\omega_1+\omega_2}{2}$, $\varepsilon \equiv \omega_1 - \omega_2$ is the frequency detuning, and $\varphi \equiv \varphi_2 - \varphi_1$ is the relative phase, whose dynamics reads

$$\dot{\varphi} = \varepsilon + \frac{\kappa}{2}\left(\frac{A_2}{A_1} - \frac{A_1}{A_2}\right)\cos(\varphi) \quad (3)$$

At steady-state, the magnitudes $A_n = A_n^S$ and the phases $\varphi_n = \varphi_n^S$ are time-independent. The solutions $(A_1^S, A_2^S, \varphi^S) \in \mathbb{R}$ of Eqs. (2,3) represent a fixed point, which we refer to as nonlinear supermodes (NLS) [40][41]. Specifically, $(A_1^S, A_2^S, \varphi^S)$ are solutions of the following coupled transcendental equations

$$\begin{cases} \gamma_1(A_1^S) \cdot \eta + \left(\frac{\kappa}{2}\right) \sin(\varphi^S) = 0 \\ \gamma_2(A_2^S) + \left(\frac{\kappa}{2}\right) \eta \sin(\varphi^S) = 0 \\ \varepsilon + \frac{\kappa}{2}\left(\frac{1}{\eta} - \eta\right)\cos(\varphi^S) = 0 \end{cases} \quad (4)$$

where $\eta \equiv A_1^S/A_2^S$, is the ratio of the field amplitudes. The nonlinear supermode frequencies $\omega$ form the RS when the Hamiltonian parameters $(\varepsilon, \kappa)$ are varied. They are given by

$$\omega = \omega_0 + \frac{\kappa}{4}\left(\frac{A_2^S}{A_1^S} + \frac{A_1^S}{A_2^S}\right)\cos(\varphi^S). \quad (5)$$

The NLS and the associated RS are categorized according to their stability which is quantified using the Lyapunov criterion. The latter requires the calculation of the eigenvalues $\{\lambda_1, \lambda_2, \lambda_3\}$ of the Jacobian matrix for the system described by Eq. (2) evaluated at the corresponding fixed points. When $\Re(\lambda_m) < 0$ for all $m = 1,2,3$, the NLS is stable, indicating that any small perturbation from the fixed-point solution will asymptotically (in time) subside in an exponential manner and the system will converge to the steady state solution. On the other hand, if $\Re(\lambda_m) > 0$ for at least one $m$, the NLS is unstable, indicating that any small perturbation from the fixed point will be exponentially amplified and the system will deviate from the steady state. In the case that $\Re(\lambda_m) = 0$, the Lyapunov criterion is not applicable, and the stability of the NLS should be characterized using alternative methods such as dynamical simulations.

### III. Nonlinear Supermodes

It is important to understand the nature of the NLS along the $\varepsilon$ and $\kappa$-directions of the Riemannian parameter space. In the $(\varepsilon = 0, \kappa)$ plane, the system described by Eqs. (1,2) supports stable fixed points whose number depends on the strength of the coupling constant $\kappa$. For strong coupling strengths $\kappa \geq 2\gamma_1(A_1^S)$, we have two stable NLS with field amplitudes $A_1^S = A_2^S = \sqrt{\gamma_1^{(0)} - \gamma_2^{(0)}/\chi_1 + \chi_2}$ and relative phase configurations satisfying $\left(\cos(\varphi_\pm^S), \sin(\varphi_\pm^S)\right) = \left(\pm\sqrt{1 - (2\gamma_1/\kappa)^2}, -2\gamma_1/\kappa\right)$ [36] [30][36]-[38]. Inserting these values into Eq. (5), we get $\omega_\pm = \omega_0 \pm \sqrt{(\kappa/2)^2 - \gamma_1^2}$ for the nonlinear supermode frequencies. In this domain, the gain in the first resonator perfectly balances the losses in the second resonator, i.e., $\gamma_1(A_1^S) = \gamma_2(A_2^S = A_1^S)$ and the steady-state Hamiltonian of Eq. (1), where $\vec{a} = (\dot{a}_1, \dot{a}_2)^T = (0,0)$, displays a self-induced parity-time ($\mathcal{PT}$) symmetry. From the above expressions for the NLS and their frequencies we conclude that the two NLS coalesce forming an NLEPD when $\kappa = \kappa_{NLEPD} \equiv 2\gamma_1(A_1^S)$.

As the coupling $\kappa$ decreases, the system transitions from the self-organized $\mathcal{PT}$-symmetric phase to a self-induced explicit $\mathcal{PT}$-symmetry breaking phase where the steady-state gain and loss components at each resonator are different, i.e., $\gamma_1(A_1^S) \neq \gamma_2(A_2^S)$. In this second configuration there is only one stable NLS. Its relative phase is $\varphi^S = -\pi/2$ resulting in $\omega = \omega_0$ (see Eq. (5)). The field amplitude at each resonator is $A_2^S = \rho A_1^S$ with $\rho < 1$ (where $\rho \in \mathbb{R}$) which reflects an energy flow from the gain resonator to the lossy one. $\rho$ can be found explicitly by solving the quartic equation $\rho^4 - \rho^3 \cdot \frac{2\gamma_1^{(0)}}{\kappa} - \rho \cdot \alpha \cdot \frac{2\gamma_2^{(0)}}{\kappa} + \alpha = 0$, where $\alpha = \frac{\chi_1}{\chi_2}$ (see Appendix I). The $\rho = 1$ value corresponds to the transition point at the NLEPD.

There are also unstable NLS whose presence affects the order of the NLEPD. For example, the Newton-Puiseux fractional series expansion of the eigenmode detuning $\Delta\omega = \sum_n c_n(\kappa, \varepsilon)\delta^{n/M}$ near the degeneracy point is controlled by the direction in the $(\varepsilon, \kappa)$-parameter space that we choose to approach the NLEPD with. By this metric, we have found two distinct fractional power law expansions. Firstly, in the $(\delta\varepsilon = 0, \kappa)$ plane $\Delta\omega \sim \delta^{1/2}$ where $\delta = \delta\kappa \equiv |\kappa - \kappa_{NLEPD}|$. This behavior characterizes EPDs of order $M = 2$. Secondly, in contrast to the $(\delta\varepsilon = 0, \kappa)$ case, along the $(\varepsilon, \kappa_{NLEPD})$ plane the dominant power in the Newton-Puiseux series is $\Delta\omega \sim \delta^{1/3}$ where $\delta = \delta\varepsilon \equiv |\varepsilon - \varepsilon_{NLEPD}|$ [36][37] (see Appendix I). Such cubic-root fractional power in the leading order term is typical for EPDs of order $M = 3$. This can be justified by the fact that Eqs. (1,2) have three NLS coalescing at the NLEPD (see Appendix I). On the other hand, the former case of the square-root fractional power in the leading order term for the $(\delta\varepsilon = 0, \kappa)$ case is rather unexpected since typically the powers that appear in the Newton-Puiseux series are integer multiples of a cubic-root. Similar peculiarities, although rare, appear also in linear EPDs of order $M = 3$ where the dominant power in the Newton-Puiseux expansion is determined by whether the perturbation respects (square root) or violates (cubic root) the parity symmetry of the structure [42].

The fractional expansion(s) of the nonlinear eigenfrequencies near the NLEPDs has been recently proposed for sensing purposes [36]-[38]. Although these proposals offer enhanced sensitivity and signal-to-noise performance, it encounters other deficiencies: it either requires a reference point for the measurements as it is not self-referenced (specifically for the cubic-root expansion $\Delta\omega \sim (\delta\varepsilon)^{1/3}$) or it assumes zero detuning $\delta\varepsilon = 0$ for the two NLS to be simultaneously stable and thus experimentally measurable. The above deficiencies of the NLEPDs-based sensing schemes implore the identification of other measurands, that maintain the enhanced sensitivity and signal-to-noise ratio, while also addressing the important issue of self-calibration. Ideally, we would also require these observables to sensitively depend on only one parameter of control in the Riemannian space while being immune to other perturbations.

### IV. Nonlinear Riemann Surfaces and Sensing Protocol based on hysteresis loop scaling

The analysis of the NLS along the ($\varepsilon = 0, \kappa$) and ($\varepsilon, \kappa_{NLEPD}$) planes give us some understanding of the structure of the Riemannian sheets, and specifically their curvatures in the proximity of the NLEPD. In the upper part of Fig. 2a, we report some typical RS associated with three different ($\chi_1, \chi_2$) configurations. In all cases, we have confirmed that the dominant fractional power in the Newton-Puiseux fractional expansion of the nonlinear supermode frequencies is directional, i.e., it depends on the specific $\kappa$ (or $\varepsilon$)-direction that we choose to perform the expansion (see black solid lines in the upper part of Fig. 2(a). The perturbation in the $\kappa$ and $\varepsilon$-direction follows a square root and cubic root dependence, respectively. Despite this "universality" there are also important structural differences between the RS for different ($\chi_1, \chi_2$) configurations.

Importantly, a bistable domain can be formed in the proximity of the NLEPD along the $\kappa > \kappa_{NLEPD}$ axis (self-induced $\mathcal{PT}$-symmetric domain) where two stable solutions coexist. In particular, the system forms a hysteresis loop. Performing a counterclockwise (CCW) path along the $\varepsilon - \omega$ plane, at a fixed $\kappa > \kappa_{NLEPD}$, the system's initial state is the $\omega_-$ stable NLS that lays on the yellow RS that occurs for very negative values of detuning, $\varepsilon = \omega_1 - \omega_2$ (see Fig. 2b). When the system's state enters the bistable region, the previous stable mode prevents other mode reaching a steady state – creating a memory effect that leads to hysteresis due to the persistence of the previous stable NLS affecting the present stable NLS. When the current mode leaves the hysteresis loop at the first critical point when $\varepsilon = \varepsilon_+$, the system abruptly switches to the $\omega_+$ stable NLS that lays on the red RS (see Fig. 2b). Likewise, as we follow the CCW path by decreasing $\varepsilon$ we enter again the hysteretic domain and at the second critical point when $\varepsilon = \varepsilon_-$ the system switches again abruptly back to the $\omega_-$ stable NLS – now completing the hysteretic loop. Its width (as measured in terms of the second Riemannian parameter $\varepsilon$) depends on the value $\delta\kappa = \kappa - \kappa_{NLEPD}$, in a power-law fashion, i.e.,

$$\Delta\varepsilon \equiv \varepsilon_+ - \varepsilon_- = (\delta\kappa)^\beta \qquad (6)$$

and it is identified by sweeping back and forth in the detuning parameter $\varepsilon$ (see turquoise arrows in Fig. 2b). We point out that similar scaling laws for the area of (dynamical) hysteresis loops versus a control parameter have been reported in the theory of dynamical systems [43][44].

We find that the exact value of the exponent $\beta$ depends drastically on the strength of the nonlinearity chosen for $\chi_1$ and $\chi_2$, i.e., $\beta = \beta(\chi_1, \chi_2)$, which controls the dynamical amplification and attenuation mechanisms involved in the system (see Appendix II). For example, when the system is linear ($\chi_1, \chi_2$) = (0,0) the RS self-intersect at the EPD without the formation of a bistable domain (see left most RS in the upper part of Fig. 2a). When $\chi_1 \neq 0$, but $\chi_2 = 0$, a bistable domain emerges, and the scaling of its width with respect to coupling, $\delta\kappa$, is superlinear as $\beta > 1$ (see lower left panel in Fig. 2(a)). Thus, the width of the hysteresis loop grows slowly for small variations in $\delta\kappa$ before dramatically blowing up. In contrast, when $\chi_1 \neq 0$, $\chi_2 \neq 0$, and $\chi_1 \approx \chi_2$, the scaling law that describes the width of the loop versus $\delta\kappa$ is sublinear, i.e., $\beta < 1$, indicating an abrupt growth with respect to the small variations in $\delta\kappa$ and consequently smaller changes in hysteresis widths and the coupling strength increases (see lower right panel in Fig. 2(a) and Appendix II).

Our investigation is guided by the TCMT analysis which predicts a sublinear scaling in Eq. (6) with $\beta \approx 0.5$, when $\chi_1 \approx \chi_2$. As the nonlinearity mismatch $\chi_1 \neq \chi_2$ increases, we find that that deviations from the $\beta = 1/2$ case occurs predominantly for small coupling variations $\delta\kappa$, see Fig. 7(a) of Appendix II. Nevertheless, the value of $\beta = 1/2$ is maintained for relatively moderate $\delta\kappa$ indicating that our scheme is robust to a small nonlinearity mismatch, i.e., $\chi_1 \approx \chi_2$.

We propose to use the sublinear scaling of the hysteresis width, $\Delta\varepsilon$, as a new measurand for hypersensitive sensing purposes. The sensing perturbation in this case is the coupling detuning, $\delta\kappa$, from the NLEPD point in $(\varepsilon, \kappa)$ parameter space – which is easily identified as the intersecting point between the cubic-root and square-root scaling of the frequency detuning that occurs when we vary $\varepsilon$ and $\kappa$ parameters, respectively. The width of the bistable domain is intrinsically self-referenced since it is estimated by the difference between the critical detuning values ($\varepsilon_\pm$) for which the mode "jumps" from one stable branch to the other. Therefore, there is no need for an external reference to suppress or eliminate frequency drifts associated with other sources. This self-reference property must be contrasted with the sensing schemes that utilize the square(cubic)-root scaling with respect to the coupling, $\kappa$ (detuning, $\varepsilon$) that always requires the prior knowledge of the position of the NLEPD point. Importantly, this sensing protocol is perturbation-specific ($\delta\kappa$-dependence), and therefore shields the sensing performance from other disturbances.

Let us finally comment on the effects of the sweeping speed of the detuning variable $\varepsilon$, on the sensitivity and precision with which an observable is estimated. We have focused our investigation to nonlinearity strengths $\chi_1 = \chi_2 = 1$ where a sublinear scaling with $\beta \approx 0.5$ in Eq. (6) is applicable. To this end, we dynamically varied the speeds of the $\varepsilon$ −sweep, while keeping $\kappa$ fixed, and extracted the width of the hysteresis loop (see Appendix II). The width of the dynamical hysteresis as a function of the coupling detuning, $\delta\kappa$, is smoothed out in the proximity of the NLEPD. Specifically, as we deviate from the adiabatic sweeping, the square-root sublinear scaling is progressively replaced by an irregular behavior (see Fig. 7(b) of the Appendix II). We conclude that there is a trade-off between sweeping time and precision sensing. The collapse of the sensing protocol as we deviate from the adiabatic sweeping, has been also reported in a recent theoretical study that proposed the use of hysteresis loop widths for sensing purposes [45] (for the universal features of the width of dynamical hysteresis in periodically switched bistable systems, see [43]). Our NLEPD-based protocol that does not require any time-consuming averaging over $\varepsilon$ −sweeping cycles to extract the hysteresis width $\Delta\varepsilon$, as opposed to the theoretical scheme of Ref. [45] which required a considerable averaging over many sweeping cycles (see also next section).

## V. Experimental Implementation

The sensing platform (see Fig. 2(c) for a schematic) is shown in Fig. 3. It consists of a pair of nonlinear $RLC$ resonators with a natural (untuned) frequency of each $LC$ tank being $f_0^{(1)} = f_0^{(2)} = f_0 = \frac{1}{2\pi}\frac{1}{\sqrt{LC}} \approx 338$ kHz ($L = 200$ μH and $C = 1110$ pF). Each $RLC$ tank $i = 1,2$ contains $R_L^{(i)}, L_i$ and $C_i$ elements which are coupled in parallel to nonlinear losses, provided by a combination of back-to-back diodes that are connected in series to a resistor $R_{NL}^{(i)}$, see Fig. 3(a). One of the $LC$ resonators incorporates a nonlinear amplifier $R_1^{-1}(V_1) = -\left(R_L^{(1)}\right)^{-1} + \left(R_{NL}^{(1)}\right)^{-1}$, which is characterized by an $I-V$ curve $I_1(V_1) = -\frac{V_1}{R_L^{(1)}} + bV_1^3$, while the other one incorporates a nonlinear loss $R_2^{-1}(V_2) = \left(R_L^{(2)}\right)^{-1} + \left(R_{NL}^{(2)}\right)^{-1}$ with an $I-V$ curve, $I_2(V_2) = \frac{V_2}{R_L^{(2)}} + bV_2^3$ ($V_1, V_2$ are the instantaneous voltage values). The nonlinearity parameter $b \approx 7 \times 10^{-4}$ AV$^{-3}$ has been chosen in a way that the width of the hysteresis loop (when sweeping the detuning parameter $\varepsilon$) varies with the coupling $\kappa$ in a square-root fashion.

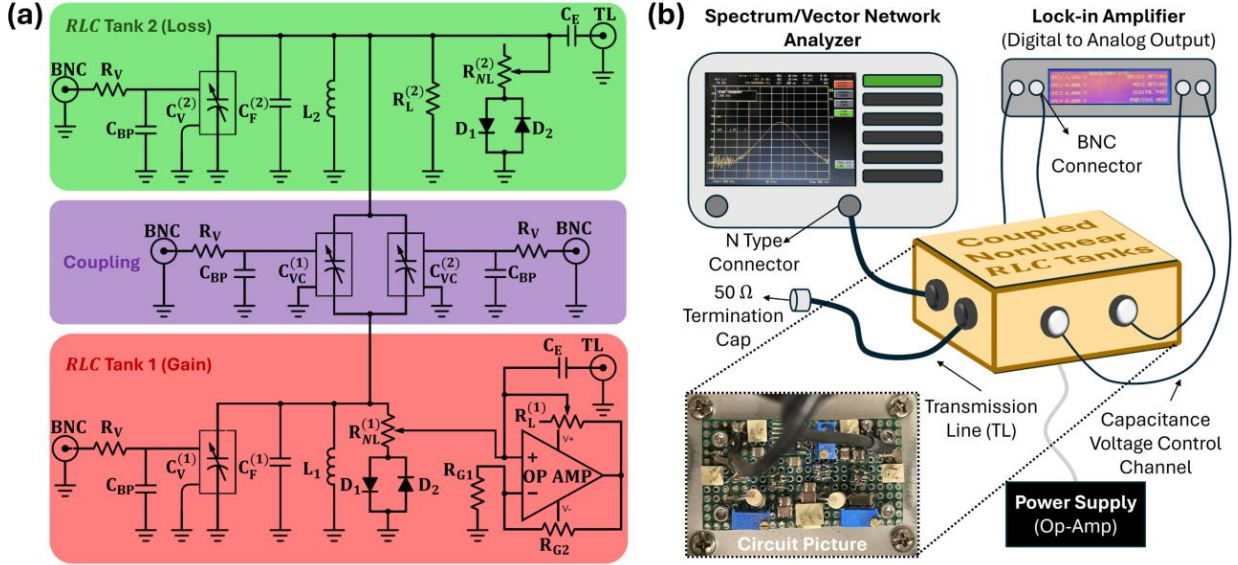

*Figure 3. Experimental circuit and measurement schematic. (a) Two RLC tanks, one with gain elements and the other with loss elements, denoted by red and green sections, respectively. The nonlinearity in each resonator is provided by the back-to-back diodes. The two RLC tanks are coupled capacitively, denoted by the purple section. (b) Overview of the connections used in the measurement. One TL port in (a) is connected to the spectrum/vector network analyzer, while the other TL has a 50 Ω termination cap. All the four BNC ports in (a) is connected to a lock-in amplifier that provides a digital-to-analog (DAC) output. These DACs allow precise control of the capacitance voltage-controlled capacitors present in the circuit. A power supply is connected to the operational amplifier (op-amp). A picture of the circuit used in the experiment is shown.*

The two $RLC$ tanks are coupled with a capacitance voltage controller (CVC) $C_\kappa(V_\kappa) = \kappa \cdot C$, where $\kappa$ is a dimensionless term that determines the strength of the coupling. These CVCs have their capacitances controlled by four BNC ports in the circuit that are connected to an EG&G Instruments 7265 DSP lock-in amplifier, see Fig. 3(b). The lock-in amplifier that has four digital-to-analog outputs (DAC) that has a resolution up to a 1 mV. Furthermore, by varying the parallel capacitance in each $LC$ tank that is parameterized as $C_\varepsilon(V_\varepsilon) = \varepsilon \cdot C$, we can control the degree of detuning between the two resonance frequencies $f_0^{(1)} - f_0^{(2)} \approx \varepsilon \cdot f_0$. The two BNC ports that control the coupling between the $RLC$ tanks $\delta V_\kappa$ are highlighted in purple in Fig. 3(a) while the two BNC ports that control the resonant frequency detuning between the $RLC$ tank $\delta V_\varepsilon$ are highlighted in the red and green sections in Fig. 3(a).

Both $\kappa$ and $\varepsilon$ are controlled by applying different voltage variations through $V_\kappa$ and $V_\varepsilon$, respectively (see Appendix III). The voltage ranges used in the experiment correspond to $1.2\text{ V} \leq V_\varepsilon \leq 3\text{ V}$ and $0.4\text{ V} \leq V_\kappa \leq 3.45\text{ V}$. In these ranges, we have confirmed that $C_{\varepsilon,\kappa}(V_{\varepsilon,\kappa})$ behave linearly with the voltage variations i.e., $C_{\varepsilon,\kappa}(V_{\varepsilon,\kappa}) \propto V_{\varepsilon,\kappa}$. The position of the NLEPD in the $(V_\varepsilon, V_\kappa)$ parameter space is identified at $V_\varepsilon^{(NLEPD)} = 2.175\text{ V}$ and $V_\kappa^{(NLEPD)} = 2.46\text{ V}$ and set to $\left(V_\varepsilon^{(NLEPD)} \to \delta V_\varepsilon,\; V_\kappa^{(NLEPD)} \to \delta V_\kappa\right) = (0,0)$ via the following relation: $\delta V_{\varepsilon,\kappa} \equiv V_{\varepsilon,\kappa}^{(NLEPD)} - V_{\varepsilon,\kappa}$. Thus, we report for voltage ranges as a function of coupling, and resonant frequency detuning away from the NLEPD in the corresponding ranges of $-0.825\text{ V} \leq \delta V_\varepsilon \leq 0.975\text{ V}$ and $-0.99\text{ V} \leq \delta V_\kappa \leq 2.06\text{ V}$. Each resonator is weakly coupled through a small capacitance $C_E = 10\text{ pF} \ll C$ to a $Z_0 = 50\text{ }\Omega$ transmission line (TL), which is used to measure the frequency spectrum of the emitted signal coming out of either resonator via a spectrum analyzer (see Fig. 3(b)). For specific details on the components used in this experiment, see Appendix III.

Since the circuit is self-oscillating (above threshold), only one transmission line (TL) is connected from either the gain or loss $RLC$ tank while the other transmission line has a $50\text{ }\Omega$ termination cap placed on it. In our experiment, the TL from the loss $RLC$ tank is connected to the spectrum/vector network analyzer, while the TL from the gain $RLC$ tank has the 50 Ω termination cap placed onto it.

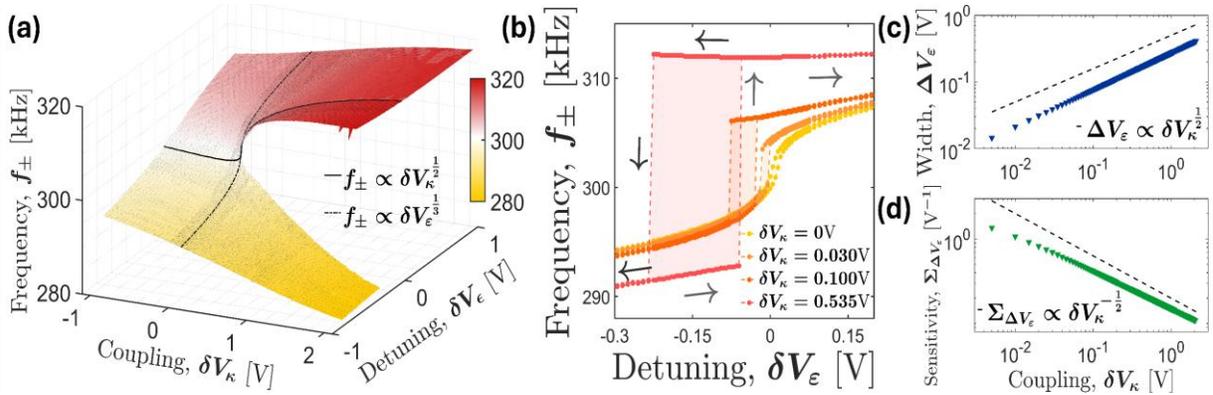

*Figure 4. **Nonlinear engineered Riemann surfaces**. In our electronic circuit platform, the coupling and detuning are controlled via voltage dependent capacitances for the coupling, $C_\kappa(\delta V_\kappa)$, and the detuning, $C_\varepsilon(\delta V_\varepsilon)$. (a) The experimental Riemann surface (corresponding to the theoretical Riemann surface with nonlinear coefficients of $\chi_1 = \chi_2 = 1$) of the coupled nonlinear RLC tank as a function of the detuning voltage, $\delta V_\varepsilon$, and the coupling voltage, $\delta V_\kappa$. The nonlinear exceptional point degeneracy (NLEPD), located at $\delta V_\kappa = \delta V_\varepsilon = 0$, demarcates the origin of the bistable domain. The supermode frequencies at $V_{\varepsilon(\kappa)} - NLEPD$ line (when $\delta V_{\varepsilon(\kappa)} = 0$), indicated by the black dashed (dashed-dotted) line, display a square(cubic)-root splitting with respect to the coupling (detuning) voltage, $f_\pm \propto \delta V_\kappa^{1/2} (\delta V_\varepsilon^{1/3})$. Qualitatively, the experimentally measured Riemann Surface agrees with the TCMT predictions (see rightmost Riemann surface in the upper part of Fig. 2(a)). (b) In our sensing protocol, the width of the hysteresis (indicative of a bistable domain) is determined by sweeping back and forth in the detuning voltage, $\delta V_\varepsilon$ (see black arrows for the downward sweep and grey arrows for the upward sweep). As the coupling voltage, $\delta V_\kappa > 0$, increases, the width of the hysteresis increases (see shaded area). (c) The experimentally measured hysteresis widths, $\Delta V_\varepsilon$, for various values of coupling voltages $\delta V_\kappa$ scale as $\Delta V_\varepsilon \propto \delta V_\kappa^{1/2}$ (a least square fit of the data gives a slightly larger power law $\Delta V_\varepsilon \propto \delta V_\kappa^{0.55}$) with the smallest measured width of 15mV. (d) The proposed sensor implementation has a sensitivity, $\Sigma_{\Delta V_\varepsilon} \equiv \frac{\partial(\Delta V_\varepsilon)}{\partial(\delta V_\kappa)} \propto \delta V_\kappa^{-1/2}$, that diverges as the NLEPD is approached at $\delta V_\kappa = 0$.*

**VI. Measurement of the Riemann Surface and Hysteresis width**

We evaluated the robustness of our platform using two different instruments, namely, a Keysight E5080A vector network analyzer, and a Rohde & Schwarz, FSEM 30 spectrum analyzer. Since the platform self-oscillates, no input signal is needed but rather only the emitted signal of the circuit is observed. To record the emitted signal, the one channel of the analyzer is connected to one of the selected output channels (either from the gain or loss $RLC$ tank) of the circuit via a weakly coupled TL while the other TL coming out of the circuit has a 50 Ω termination cap placed onto it (see Figs. 3(a,b)). Regardless of the use of the spectrum analyzer or the vector network analyzer, both pieces of instrumentation collect the Fourier spectrum of the emitted signal from the circuit. To measure the Riemann surface, the individual frequency sweeps were conducted: on the Keysight E5080A in a range of $(275 - 325)$ kHz with 501 points and an intermediate frequency bandwidth (IFBW) of 500 Hz; on the Rohde & Schwarz FSEM 30 in the same range of $(270 - 325)$ kHz but with 500 points and a radio and video bandwidth (RBW & VBW) of 500 Hz.

The measured frequencies $f_\pm$ of the experimental RS are shown in Fig. 4(a). To capture the bistable domain, $\delta V_\varepsilon$ was adiabatically swept back and forth while $\delta V_\kappa$ was held constant, see Fig. 4(b). Using this protocol, the width of the hysteresis, $\Delta V_\varepsilon = \delta V_\varepsilon^{(f+)} - \delta V_\varepsilon^{(f-)}$, was extracted by determining extreme values of $\delta V_\varepsilon$ at which the NLS frequency jumps from the $f_-$ to the $f_+$ NLS, see Fig. 4(b). It is important to stress that the shape of the hysteretic loop is reproducible after any sweeping cycle – a feature that allows us to avoid a time-consuming averaging process for extracting the hysteresis width $\Delta V_\varepsilon$ for each value of $\delta V_\kappa$, i.e., a single measurement is sufficient (compare with Ref. [45] where a substantial averaging was necessary for realizing a sublinear scaling). The measured $\Delta V_\varepsilon$ values for several different voltages $\delta V_\kappa > 0$ is reported in Fig. 4(c), following closely a square root scaling law $\Delta V_\varepsilon \propto \delta V_\kappa^{1/2}$. These experimental results are in agreement with the predictions of TCMT, which has been used to optimize the choice of the nonlinear parameter $b$ (see also Appendix II). From the $\Delta V_\varepsilon(\delta V_\kappa)$ curve, the sensitivity, $\Sigma_{\Delta V_\varepsilon}$, is found by taking the derivative, $\Sigma_{\Delta V_\varepsilon} = \frac{\partial(\Delta V_\varepsilon)}{\partial(\delta V_\kappa)}$, and it confirms the TCMT prediction as it scales as $\Sigma_{\Delta V_\varepsilon} \propto \delta V_\kappa^{-\frac{1}{2}}$ (see Fig. 4(d)). The diverging behavior of the sensitivity in the proximity of the NLEPD guarantees a dramatically enhanced transduction operation that maps an input parameter to an output measurable quantity.

**VII. Noise Analysis**

Conceptually, sensing can be broken down into two components: the portion that directly measures the quantity of interest due to deterministic processes (the signal) and the measurement distortion due to noise. We have already shown that an appropriate design of the nonlinearities can optimize the first component. However, it is vital to study both components *in-tandem*. Especially for self-oscillating sensing schemes, there are various noise sources associated with the amplification mechanisms utilized for the realization of NLEPDs. Obviously, the sensor efficiency has to account for its resilience against noise, which in some cases has hindered, or even limited, the measurement of the small variations of the perturbation parameter in case of linear EPDs [11][46]-[48].

To analyze the effect of noise present in the evaluation of the width of the bistable loop $\Delta V_\varepsilon$, we utilize the Allan deviation $\sigma_{\widetilde{\Delta V_\varepsilon}}(\tau)$ which essentially quantifies the fluctuations of the $\Delta V_\varepsilon$ −measurements with respect to an expected value at a specific sampling time $\tau$. It is defined as [49]-[51],

$$\sigma_{\widetilde{\Delta V_\varepsilon}}(\tau) = \sqrt{\frac{1}{2(K-1)} \sum_{n=1}^{K-1} \left(\langle \widetilde{\Delta V_\varepsilon}^{(n+1)}(\tau) - \widetilde{\Delta V_\varepsilon}^{(n)}(\tau) \rangle\right)^2} \qquad (7)$$

where $\Delta V_\varepsilon$ is normalized with respect to the experimental location of the NLEPD, $V_\kappa^{(NLEPD)}$, such that $\Delta \widetilde{V_\varepsilon} = \Delta V_\varepsilon / V_\kappa^{(NLEPD)}$. Above, $\tau \equiv K\tau_0$ indicates the duration of the sampling time where $\tau_0$ is the minimum measurement time needed to sweep back-and-forth in $\delta V_\varepsilon$ to acquire the width of a single hysteresis loop $\Delta V_\varepsilon$ for a specified $\delta V_\kappa$, $K \leq N/2$ is a positive integer that denotes the size of a sampling cluster, and $N$ indicates the number of data acquisitions. In our case, the measurement of the bistable region $\Delta V_\varepsilon$ was repeated for a total of $N = 500$ acquisitions in a consecutive fashion for a total time $t \approx [35, 35, 34, 35, 35, 56, 58] \times 10^3$s for $\delta V_\kappa = [0.005, 0.01, 0.02, 0.055, 0.105, 0.350, 0.650]$ V away from the NLEPD and $\tau_0 \approx [69, 69, 67, 69, 69, 112, 115]$ s.

The noise analysis conducted theoretically and experimentally is reported in Fig. 5. In Figs. 5(a, b) we report the Allan deviation calculated from (a) TCMT (as $\sigma_{\Delta\tilde{\varepsilon}}$) and (b) experimentally (as $\sigma_{\Delta\widetilde{V_\varepsilon}}$) for various representative values of coupling perturbations $\delta V_\kappa$ that span the range of being in proximity to and far away from the NLEPD. The TCMT modeling in Fig. 5(a) shows that the noise level remains roughly the same (or even is reduced as the NLEPD is approached!) irrespective of the value of $\delta V_\kappa$. These calculations are corroborated experimentally as the behavior of $\sigma$ in Fig. 5(b) indicates similar trend. In fact, using the TCMT modeling, we were able to extend the noise analysis to smaller sampling times $\tau$ (compared to the experiment) and confirm that the above results remain unchanged. To better quantify the effects of noise in the measurement process we have also analyzed the noise-to-signal ratio described by the normalized Allan deviation, $\sigma_\alpha \equiv \frac{\sigma_{\Delta V_\varepsilon}}{\Sigma(\delta V_\kappa)}$, in Figs. 5(c, d). Both the TCMT simulations (in (c) as $\sigma_\alpha^{\Delta\tilde{\varepsilon}}$) and the experimental measurements (in (d) as $\sigma_\alpha^{\Delta\widetilde{V_\varepsilon}}$) confirm that the normalized Allan deviation decreases as the coupling between the two $RLC$ tanks approach the NLEPD at $\delta V_\kappa = 0$. In this domain of small perturbations, the sensitivity diverges and given that the noise is (roughly) constant, noise-to-signal suppression is observed – boosting the overall sensing performance. In fact, the observed noise-to-signal suppression is comparable to the sensing protocol that utilizes nonlinear frequency detuning (for $\delta \varepsilon = 0$) in the proximity of the NLEPD [36] while it is more than an order of magnitude larger than the suppression obtained using linear EPD sensing schemes [8].

Further analysis on the dependence of the Allan deviation of the experimental data (see Figs. 5(b,d)) with respect to the sampling time $\tau$ leads to the identification of various sources of noise that affect the performance of the sensor [50]. For long sampling times $\tau$, the Allan deviation shows a behavior $\sigma_\alpha \propto \tau^{1/2}$, indicating that the circuit is affected by a rate random walk (RRW), a process characterized by an exponentially correlated fluctuation whose correlation time $\tau_c$ is much larger than a cluster time $\tau \equiv K\tau_0$ [49]. During intermediate sampling times $\tau$, the Allan deviation saturates at a certain value $\sigma_\alpha = \alpha_{BI} \cdot \tau^0$ suggesting the presence of bias instability (BI) noise. The coefficient $\alpha_{BI}$ determines the smallest possible measurement of the bias present within our sensor. This noise stems from low-frequency fluctuations present in either electronic components or other parts of the system that are susceptible to random flickering. Finally, the short-time behavior of the noise is associated to voltage random walk (VRW) behavior characterized by an Allan deviation that scales as an inverse square root with respect to the sampling time $\tau$, as $\sigma_\alpha = \alpha_{VRW}\tau^{-1/2}$. The source of this noise is associated with the high-frequency fluctuations that affect the sensor, like thermal (Johnson-Nyquist) noise inducing random voltage fluctuations in the resistive elements in the circuit and/or in the signals controlling the capacitor values; readout noise; or other noise sources with correlation time much shorter than the measurement time (white-noise). The contributions of all different sources add up into an effective voltage "random walk" whose magnitude relative to the sensitivity is encoded in a single parameter $\alpha_{VRW}(\delta V_\kappa)$. We point out that the TCMT, addresses only the impact of VRW noise which features only a $\tau^{-1/2}$ decay of the Allan deviation that occur at short sampling times which is the most relevant for sensing purposes (see Figs. 5(a,c)). From TCMT simulations (see Figs. 4(a,c)) and experimental measurements (see Figs. 5(b,d)), we extract the values of $\alpha_{VRW}$ for various voltage variations, $\delta V_\kappa$ (see Figs. 5(e,f)). According to the experiment (f), the line of best fit gives $\alpha_{VRW} \approx 0.084 \cdot \delta V_\kappa^{0.51}$. Over the entire range of $\delta V_\kappa$, there is over one-order of magnitude (~15) suppression in the $\alpha_{VRW}$ noise coefficient, a clear manifestation of the resilience of the sensing protocol against noise. Such a noise suppression is also confirmed by the TCMT model (e) – albeit here it is even more pronounced indicating a higher power divergency in the vicinity of the NLEPD, as $\alpha_{VRW} \approx 0.02 \cdot \delta V_\kappa^{0.95}$. This is to be expected as the TCMT modeling accounts only for the Johnson-Nyquist noise and does not consider other noise sources that affect the performance of the sensor.

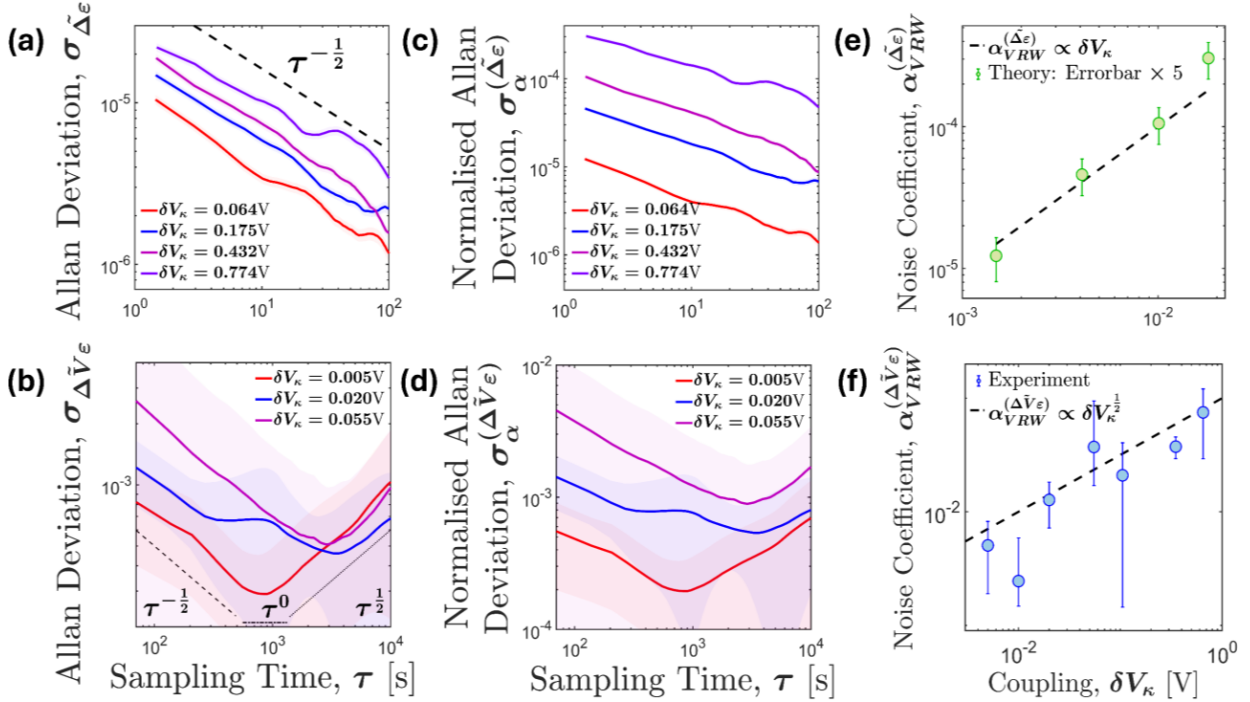

*Figure 5. Noise analysis. Allan deviation $\sigma$ of the rescaled hysteresis width as a function of sampling time $\tau$, for different values of the coupling – reported as $\delta V_\kappa$: (a) using temporal coupled mode theory (TCMT) for the rescaled (numerically extracted) hysteresis width $\widetilde{\Delta\varepsilon} \equiv \Delta\varepsilon/\kappa_{NLEPD}$; (b) experimental results for the rescaled measured hysteresis width $\widetilde{\Delta V_\varepsilon} \equiv \Delta V_\varepsilon/V_\kappa^{(NLEPD)}$. The TCMT coupling constant indicated in (a) is associated with the experimental coupling voltage $\delta V_\kappa$ via the relation $\delta\kappa = 0.0234\,\delta V_\kappa$. The numerical simulations have been extrapolated to small sampling times in order to confirm the observed experimental behavior of Allan deviation beyond the minimum experimental sampling time limitations. The noise level indicated by the Allan deviation $\sigma$ does not change significantly as the coupling approaches the NLEPD in both (a, b). A black dashed line indicates a $\tau^{-1/2}$ behavior in the short time – a signature of the phase random walk noise of the circuit. In (b) we also indicate by a dashed-dotted line in the intermediate sampling time domain where bias instabilities ($\sigma \sim \tau^0$) dominate and by a dotted line in the long sampling time domain of denoting the rate random walk ($\sigma \sim \tau^{1/2}$) noise processes. The noise-to-signal performance of our sensor is better characterized through the normalized Allan deviation, $\sigma_\alpha \equiv \frac{\sigma}{\Sigma}$, where $\Sigma$ is the sensitivity for (c) the TCMT, and (d) the experiment. As the nonlinear exceptional point degeneracy (NLEPD) is approached by decreasing the coupling $\delta V_\kappa$, the signal sensitivity overcomes the noise, resulting in reduced values of $\sigma_\alpha$. Shaded region in (a, b, c, d) represent an error of $\pm 1$ standard deviation. (e, f) The extracted values of $\alpha_{VRW}$ in the short sampling times where $\sigma_\alpha = \alpha_{VRW}\tau^{-1/2}$ for various voltage couplings. The TCMT in (e) indicates a $\alpha_{VRW} \propto \delta V_\kappa$ scaling (black dashed line) while the experimental measurements in (f), indicate a much slower scaling $\alpha_{VRW} \propto \delta V_\kappa^{1/2}$ (black dashed line), due to the presence of other noise sources that are not accounted for in the TCMT modeling.*

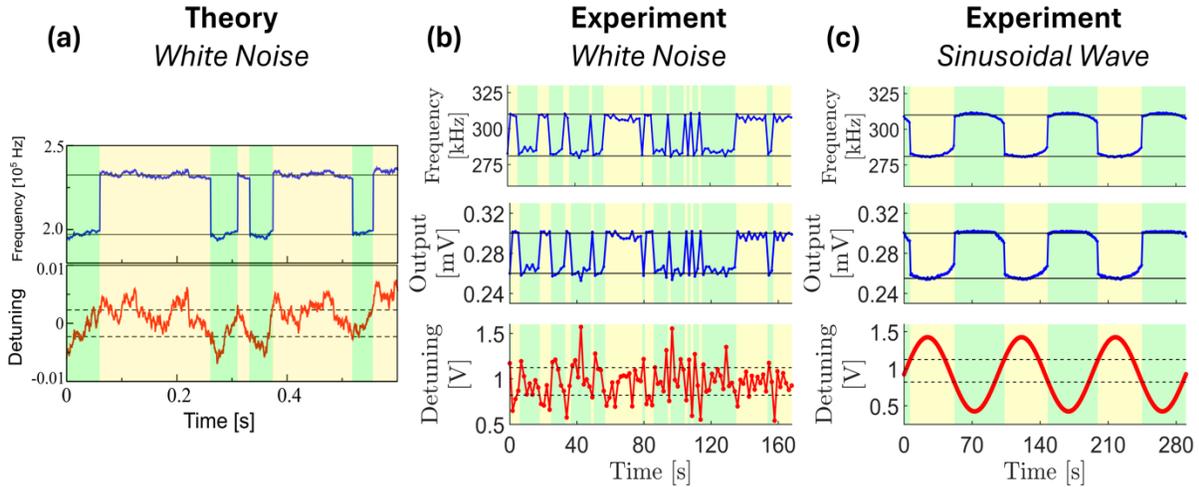

*Figure 6. Schmitt trigger operation. (a) A theoretical simulation using TCMT demonstrates that the frequency takes two approximately constant values (upper panel) – digital output signal – (see horizontal solid lines) when the detuning undergoes a random walk (lower panel) –analog input signal. The jumps between frequency-values occur at the edges of the bistable domain (see horizontal dashed lines). (b) Experimental data from our platform in which the detuning varies as white noise following a voltage random walk (lower panel). The output frequency and voltage with respect to time is shown in the upper and middle panel, respectively. (c) As in (b), but now the detuning is varied in a sinusoidal manner. The light green (yellow) boxes denote the time-domains during which the system has switched to the $f_+(f_-)$ nonlinear supermodes.*

## VIII. Implementation of Schmitt Triggers

The hysteresis formed in the proximity of the NLEPD can be also utilized for developing a Schmitt trigger. A Schmitt trigger is a device (traditionally an electronic circuit), which converts an analog signal into a digital output. It is particularly useful in handling noisy signals as the presence of hysteresis enforces well defined threshold voltages for stabilizing the output signal between two values. Given that our platform supports such a controlled hysteresis loop, we examine its viability for such usage.

Figure 6(a) represents the results of TCMT simulations whilst Fig. 6(b-c) represents the experimental results for two types of varying analog input signals. In Fig. 6(a), we consider that the frequency detuning between the two RLC tanks is dictated by a white noise process. By exploiting our setup as a Schmitt trigger, we observe that whenever the detuning crosses the edges of the bistable domain (see horizontal dashed lines in Fig. 6(a)), the frequency jumps between the two values associated with the stable nonlinear supermodes that determine the Riemann surface. The edge-values of the bistable domain define the threshold values that dictate the frequency of the output signal of the Schmitt trigger. The in-between values of the detuning do not "contaminate" the nonlinear supermode frequency of the output signal which acquires only the two threshold frequency values. Thus, our proposed Schmitt trigger delivers a binary (digital) output signal characterized only by the two stable nonlinear supermode frequencies $f_\pm$ (see Fig. 6(a)). The small variation in the frequency of the output signal is associated with the non-flat topology of the Riemann surfaces when varying $\varepsilon$. However, for all practical purposes the output signal can be treated as binary.

In Fig. 6(b), we present experimental results of the proposed Schmitt trigger for the case that the detuning voltage, $\delta V_\varepsilon$, is altered in time as white noise (following a random walk process—see lower panel of Fig. 6(b). In the other two panels, we are reporting the voltage (middle) and frequency (upper) extracted from the measured power spectrum of the output signal. The coupling is set to $\delta V_\kappa = 2.06V$. Initially, the system is in the $f_-$ NLS. As the detuning voltage that controls the natural frequencies of the two RLC tanks crosses the upper threshold (see upper horizontal dashed line in the lower panel of Fig. 6(b)), the frequency jumps from $f_-$ NLS to $f_+$ NLS. The frequency of the platform remains unchanged even as it crosses the upper threshold voltage once more, and only changes back to the $f_-$ NLS upon crossing the lower threshold voltage (see lower horizontal dashed line in the lower panel of Fig. 6(b)). Thus, the gap between the upper and lower detuning voltage thresholds in the lower panel (see gap between horizontal dashed lines) represents the noise buffer. This buffer enables our platform to trigger only upon sufficiently large changes in the detuning voltage for the system to flip-flop between the $f_\pm$ NLS.

Similar experimental results are shown in Fig. 6(c), for the case of a sinusoidal detuning voltage $\delta V_\varepsilon$. Despite the different analog input signals, our proposed Schmitt trigger delivers an output signal with the same binary voltage and frequency values as in the previous case of Fig. 6(b).

## Discussion

We have experimentally and theoretically demonstrated a new noise-resilient, high-precision sensing scheme. The sensing scheme utilizes the existence of bistable domains of the Riemann Surfaces (RS) of nonlinear supermodes (NLS) of a self-oscillating system consisting of two $RLC$ coupled nonlinear resonators with saturable amplification and attenuation mechanisms. For appropriate values of the RS parameters (coupling $\kappa$, resonant frequency detuning $\varepsilon$), this system supports a NLEPD which demarcates the birth of a bistable domain that supports the formation of a hysteresis loop as sweeps in $\varepsilon$ are made. The width of this loop scales in a power-law fashion with the coupling detuning $\delta\kappa$ away from the NLEPD value of $\delta\kappa = 0$. The value of the power law exponent (and therefore, the degree of sensitivity) is controlled by the strengths of the nonlinear coefficients, $\chi_{1,2}$, which act as a design knob. We have identified appropriate nonlinear configurations $\chi_{1,2}$ that allowed us to engineer a square-root scaling of the hysteresis width that displays an enhanced sensitivity to small coupling variations. The upper bound of this extraordinary sensing precision is controlled by the speed of the $\varepsilon$- sweeps and becomes optimal in the adiabatic case.

As opposed to other NLEPD schemes, the current proposal does not need fine tuning to adjust for the precise NLEPD point and is intrinsically self-referenced, i.e., there is no need for an external reference to suppress or eliminate frequency drifts associated with other sources since it measures the difference between the minimum and maximum values of $\varepsilon$ that mark the edges of the hysteresis loop. Importantly, the sensitivity and sensing precision of the signal in the case of small coupling variations remains unaffected by the various noise sources which maintain the same (or even reduced!) level of influence for all coupling variations (close and far from the NLEPD). As a result, in the vicinity of NLEPD (small coupling variations), the noise-to-signal ratio is dominated by the extraordinary signal sensitivity of the proposed sensing protocol.

Our prototype experiment demonstrates that the nonlinearity-aided RS and hysteresis engineering near NLEPDs for self-referential, noise-resilient sensing can outperform linear EPD sensors. Our sensing scheme does not require any averaging over many sweeping cycles for the evaluation of the hysteresis loop and it can be realized in simple platforms like the one shown in Fig. 2(c). In fact, hysteresis loop engineering can offer other modalities. Along these lines, one can envision its use for $\kappa$ −reconfigurable Schmitt triggers and bistable multivibrators (latches and flip-flops). The interplay of non-Hermiticity and nonlinearities opens up new research directions that go beyond the sensing framework – covering areas like wavefront shaping [52], frequency combs [53], optical signal processing and nonlinearity-protected memory devices [54][53][55].

**Acknowledgements**

A.S., M.R., and T.K. acknowledges their partial support from MPS Simons Collaboration via grants No. 733698 and No. SFI-MPS-EWP-00008530-08 and grant NSF ECCS 2148318 – the last of which is supported in part by funds from OUSD R&E, NIST, and industry partners as specified in the Resilient & Intelligent NextG Systems (RINGS) program. U.K. and L.F.-A acknowledge MPS Simons Collaboration grant No. 733698 which partially supported their visit to Wesleyan University. L.F.-A., and P.F.W.-B. acknowledge their partial support from CONICET and MINCyT Grant No. CONVE-2023- 10189190—FFFLASH. A.S. acknowledges Mr. Miroslaw Koziol discussions and assistance in fabrication of the circuit.

# APPENDICES

## Appendix I: Nonlinear Supermodes

Here, we provide details about the analytical derivation of the nonlinear supermodes (NLS) and their frequencies along the $\varepsilon$ and $\kappa$-directions of the Riemannian parameter space discussed in Sec. III.

We start by discussing the case of $(\varepsilon = 0, \kappa)$, i.e., when there is no frequency detuning between the resonators. To satisfy Eq. (3) two scenarios are possible: (i) the phases of the fields are orthogonal, i.e., $\cos(\varphi^S) = 0$; or (ii) the magnitudes of the fields are the same, $A_1^S = A_2^S$.

In the first scenario, where $\cos(\varphi) = 0$, i.e., $\varphi = \pm\frac{\pi}{2}$, the NLS are characterized by a frequency $\omega = \omega_0$. From the first two sets of Eq. (2), we conclude that only the relative phase $\varphi^S = -\frac{\pi}{2}$ ensures solutions with $\gamma_{1(2)}, \left(A_{1(2)}^S\right)^2 \geq 0$. In this case, the NLS can have different field amplitudes, i.e., $A_1^S \neq A_2^S$ and the corresponding field intensities at each resonator are obtained from $\frac{2\gamma_1}{\rho} = \kappa$ and $2\gamma_2\rho = \kappa$, via the explicit functional form of $\gamma_{1(2)}$. Here, $\rho \equiv \frac{A_2^S}{A_1^S}$, represents the relative field amplitude, and, in the case of $\gamma_n = \gamma_n(|A_n|^2) = \gamma_1^{(0)} + (-1)^n \chi_n |A_n|^2$ analyzed in Sec. III, is obtained from the quartic equation

$$\rho^4 - \rho^3 \cdot \frac{2\gamma_1^{(0)}}{\kappa} - \rho \cdot \alpha \cdot \frac{2\gamma_2^{(0)}}{\kappa} + \alpha = 0 \tag{8}$$

where $\alpha = \frac{\chi_1}{\chi_2}$. Out of the four solutions of the quartic Eq. (8) one has to select the ones satisfying (a) $\rho \in \mathcal{R}_{>0}$, (b) $\rho \leq \frac{2\gamma_1^{(0)}}{\kappa}$ such that $(A_1^S)^2 \geq 0$, and (c) $\rho \leq \frac{\kappa}{2\gamma_2^{(0)}}$ such that $(A_2^S)^2 \geq 0$. It turns out that only $\rho \leq 1$ provides stable NLS solutions, and it is consistent with our intuition that energy leaks from the gain to the lossy resonator.

In the second scenario, where the field intensities are uniform within the dimer, i.e., $A_1^S = A_2^S$, the system arrives to a self-organized PT-symmetric configuration where the gain is perfectly balanced by the loss, $\gamma_1(A_1^S) = \gamma_2(A_2^S)$. In the case of the nonlinear functions $\gamma_1(A_1) = \gamma_1^{(0)} - \chi_1 A_1^2$ and $\gamma_2(A_2) = \gamma_2^{(0)} + \chi_2 A_2^2$, the field intensities become $A_1^S = A_2^S = \sqrt{\frac{\gamma_1^{(0)} - \gamma_2^{(0)}}{\chi_1 + \chi_2}}$. This domain is restricted to relatively strong coupling strengths, $\kappa \geq 2\gamma_1$, where there is a coexistence of two NLS where the fields in the resonators are able to self-organize in one of two possible configurations with fixed relative phases satisfying $(\cos(\varphi_\pm), \sin(\varphi_\pm)) = \left(\pm\sqrt{1 - \left(\frac{2\gamma_1}{\kappa}\right)^2}, -\frac{2\gamma_1}{\kappa}\right)$ From these expressions, we conclude that two NLS coalesce forming a NLEPD when $\kappa = \kappa_{NLEPD} \equiv 2\gamma_1(A_1^S)$. The presence of the NLEPD is further confirmed when analyzing the NLS' frequencies $\omega_\pm = \omega_0 \pm \sqrt{\left(\frac{\kappa}{2}\right)^2 - \gamma_1^2}$ associated to those solutions where $\omega_{NLEPD} = \omega_0$.

Interestingly, the direction in the Riemannian parameter space that we choose to approach/depart from the NLEPD determines the fractional-power law associated to the splitting of the frequencies. While perturbations in the coupling, $\kappa = \kappa_{NLEPD} + p$, yield a $\delta\omega \sim p^{1/2}$, perturbations in the detuning, $\varepsilon \neq 0$, can generate a higher order root frequency splitting $\delta\omega \sim \varepsilon^{1/3}$. The latter case can be shown analytically by considering the scenario of a single nonlinear gain saturation and linear loss, i.e., $\chi_1 = 1, \chi_2 = 0$ [30][37].

To this end, we move forward by solving the eigenvalue equation from Eq. (1), by assuming the existence of a NLS with frequency $\omega$. Solutions for $\omega$ are obtained from the resulting secular equation,

$$(\omega_1 - \omega)(\omega_2 - \omega) + i[\gamma_1(\omega_2 - \omega) - \gamma_2(\omega_1 - \omega)] + \gamma_1\gamma_2 - \kappa^2 = 0 \qquad (9)$$

To ensure the existence of a real solution, the nonlinear gain and loss elements are required to satisfy $\gamma_1(\omega_2 - \omega) = \gamma_2(\omega_1 - \omega)$, thus, eliminating the complex coefficient term. Without loss of generality, we consider $\omega_2 = \omega_0$ and $\omega_1 = \omega_0 + \varepsilon$, where $\varepsilon$ is the detuning. Then, Eq. (9) can be rewritten as,

$$(\omega - \omega_0)^3 - \varepsilon(\omega - \omega_0)^2 - (k^2 - \gamma_2^2)(\omega - \omega_0) - \gamma_2^2 \varepsilon = 0 \qquad (10)$$

When $\varepsilon = 0$, the three solutions contain $\omega = \omega_0$, besides the expected $\omega_\pm = \omega_0 \pm \sqrt{(\kappa/2)^2 - \gamma_2^2}$, which feature the NLEPD at $\kappa_{NLEPD} = 2\gamma_2 = 2\gamma_1$ discussed in the previous section.

Now, we explore the parameter space through the $(\varepsilon, \kappa_{NLEPD})$ plane, where the linear term in Eq. (10) vanishes. The solution to this equation is obtain by using Cardano's formula,

$$\begin{cases} w_1 = \dfrac{1}{3}\left(e - C(e) - \dfrac{e^2}{C(e)}\right) \\ w_{2(3)} = \dfrac{1}{3}\left(e - \dfrac{-1 \pm i\sqrt{3}}{2}C(e) - \dfrac{2e^2}{(-1 \pm i\sqrt{3})C(e)}\right) \end{cases} \qquad (11)$$

where $w = (\omega - \omega_0)/\gamma_2$, $e = \varepsilon/\gamma_2$, and $C(e) = \sqrt[3]{-e^3 - (27e \pm |e| \cdot \sqrt{729 + 108e^2})/2}$. In the limiting case of $e \ll 1$, $C(e) \approx \alpha_\pm \sqrt[3]{e}$, with $\alpha_\pm = \sqrt[3]{-(27 \pm \sqrt{729})/2}$, and therefore

$$\begin{cases} w_1 \approx -\dfrac{1}{3}C(e) \propto \sqrt[3]{e} \\ w_{2(3)} \approx -\dfrac{-1 \pm i\sqrt{3}}{6}C(e) \propto \sqrt[3]{e} \end{cases} \qquad (12)$$

thus, making apparent the cubic root splitting in the vicinity of the NLEPD, which requires three coalescing NLS. We have also verified via detailed numerical simulations that this cubic splitting scenario remains valid for other cases where $\chi_2 \neq 0$.

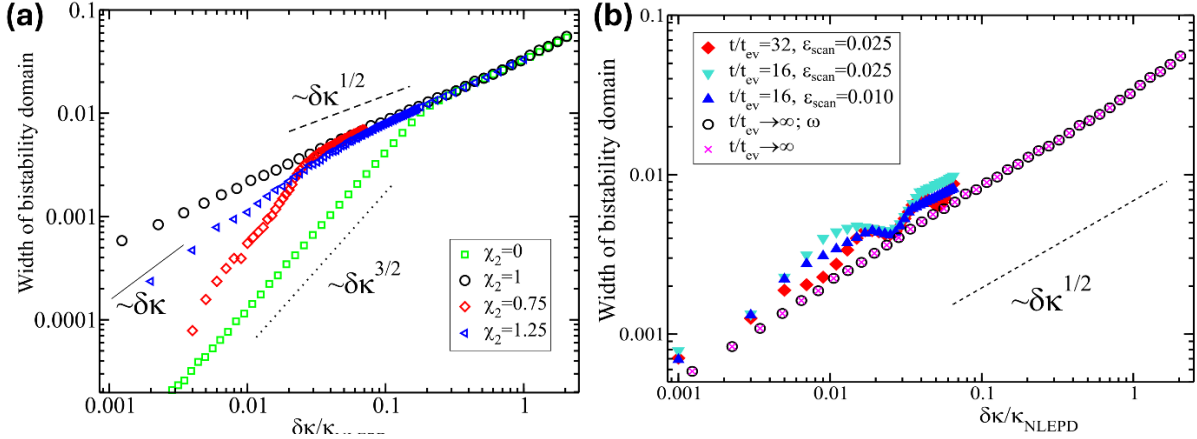

*Figure 7. Dynamical exploration of the bistable domain. We explore the Riemann surfaces by sweeping in the detuning $\varepsilon$ while keeping $\kappa = constant$. (a) We report the width of the bistable domain $\Delta\varepsilon = \varepsilon_+ - \varepsilon_-$, for different nonlinear $\chi_2$-configurations as extracted using dynamical simulations with extremely slow sweeping speed (adiabatic sweeping). The width was evaluated by identifying the NLS frequencies $\omega_\pm$. When $\chi_2 = \chi_1(=1)$ a persistent square-root scaling of the hysteresis width is maintained over more than three orders of magnitude versus coupling detuning from NLEPD. (b) When increasing the sweeping speed (reducing the evolution time $t_{ev}$) of the optimal nonlinear configuration $\chi_2 = \chi_1(=1)$, the width of the bistable domain acquires an irregular scaling behavior. Such behavior also occurs when varying the $\varepsilon_{scan}$-exploration amplitude, see dark blue triangles. The evaluation of the hysteresis domain in (b) was done by identifying jumps in the amplitudes $A_1$. We have checked that for extremely slow (adiabatic) sweeping the extracted hysteresis width using $A_1$ is identical to the hysteresis width by analyzing the NLS frequencies $\omega_\pm$ (open black circles). Dashed, dotted and solid black lines are drawn to guide the eye and indicate powers $\frac{1}{2}, \frac{3}{2}$ and $1$, respectively.*

## Appendix II: Dynamically Exploring the Bistable Domain by Varying the Detuning

In this section we provide details about the optimal nonlinear configuration and the dynamical exploration of the bistable domain that originates the hysteresis loop. Our analysis relies on the TCMT which allows us to engineer an optimal hypersensitive sensing platform. To this end, we solve Eq. (1) numerically using a Runge-Kutta (RK4) numerical integrator, keeping $\kappa$ fixed and sweeping back and forth in the detuning for various $\chi_2$-values while $\chi_1 = 1$ is kept fixed. To appropriately determine the width of the bistable domain, we vary the detuning in $\varepsilon$-steps following the prescription $\varepsilon_n = \varepsilon_{scan} \cdot \cos(2\pi n/N_\varepsilon)$, with $N_\varepsilon \sim 2000$, and letting the system evolve during a time $t_{ev}$ keeping $\varepsilon_n = constant$ before evaluating any of the dynamical variables $A_1, A_2$ or $\varphi$ and then making the next $\varepsilon$-step. While $\varepsilon_{scan}/N_\varepsilon$ controls the $\varepsilon$-step size, the evolution time, $t_{ev}$, is a control variable that determines whether an evolution of the sweep is adiabatic or fast when compared with the oscillation period of the bare resonator, $t_0 \equiv 2\pi/\omega_0$. The amplitude of the sweeping, $\varepsilon_{scan}$, must be large enough so to produce jumps to the only stable Riemann sheet at each edge of the bistable domain.

We start by analyzing the hysteresis loop obtained through an adiabatic scan ($t_{ev} \geq 300 \cdot t_0$) of the detuning $\varepsilon$ for various nonlinear configurations i.e. varying $\chi_2$ while keeping $\chi_1 = 1$ constant, see Fig. (7). As discussed in Sec. III, for $\kappa > \kappa_{NLEPD}$, it is possible to find two stable solutions with associated frequencies $\omega_\pm$, whose corresponding splitting $\Delta\omega = \omega_+ - \omega_-$ is different than zero only when the two $\omega_\pm -$ solutions exist. In the (quasi-)adiabatic case, we use the frequency splitting to analyze the width of the bistable domain. However, the same results (as far as the scaling of the hysteresis width is concerned) can be also obtained by monitoring the steady state values of $A_1, A_2$ or $\varphi$ (see, for example, the magenta crosses in Fig. 7(b) that are using $A_1$ for the evaluation of the hysteresis width and compare with the open black circles that use $\omega_\pm$). We find that for small coupling variations $\delta\kappa$, the width of the bistable domain grows with different powers $\Delta\varepsilon \propto \delta\kappa^\beta$ depending on the value of $\chi_2$, see Fig. 7(a). Nevertheless, the value of $\beta = 1/2$ is maintained for relatively moderate $\delta\kappa$ indicating that our scheme is robust to a small nonlinearity mismatch, i.e., $\chi_1 \approx \chi_2$. When $\chi_2 = \chi_1(= 1)$ a persistent square-root scaling of the hysteresis width $\Delta\varepsilon \sim \delta\kappa^{1/2}$ is maintained over more than three orders of magnitude versus coupling detuning from the NLEPD $\delta\kappa = \kappa - (\kappa_{NLEPD} \pm \sigma_\kappa)$, where $\sigma_\kappa \ll \kappa_{NLEPD}$ is a very small correction to the coupling perturbation from the NLEPD that is raised due the quasi-adiabatic nature of the simulations, specific form of driving etc. This behavior has to be contrasted to other $\chi_2$ −cases where other power laws with larger exponent $\beta > 0.5$ are observed for small coupling detunings. Such configurations do not present any advantage for hypersensitive sensing.

Next, we focus our numerical analysis to the optimal case $\chi_1 = \chi_2 = 1$ and analyze the effects of dynamical sweeping in the scaling of the hysteresis. In this case, the evaluation of the hysteresis width has been done by identifying the $\varepsilon_\pm$-values for which $A_1$ becomes bistable. Here, not only the characteristic time, $t_{ev}$, controls the dynamical exploration of the loop but also the strength of the detuning steps $\sim \varepsilon_{scan}/N_\varepsilon$. As captured by our numerical simulations Fig. 7(b), when $t_{ev}$ decreases, the scaling of the width of the bistable domain becomes irregular; hence, affecting the adiabatic scaling behavior $\Delta\varepsilon \propto \delta\kappa^{1/2}$.

**Appendix III: Circuit Design and Measurement Set-Up**

The experimental setup (see Fig. 3(a)) consists in two $RLC$ tanks, coupled through voltage-controlled capacitors, $C_\kappa = \kappa \cdot C \equiv (C_{VC}^{(1)} + C_{VC}^{(2)})$. The inductors in both $RLC$ tanks are identical (API Delevan 807-1537-90HTR) with $L_i = 200$ µH where $i = 1$ (2) denotes the $RLC$ tank with gain (loss). The effective capacitance $C_i$ in each resonator consists of two parallel elements $C_F^{(i)} = 910$ pF (Kemet 80-C0603C911F5G) and $C_V^{(i)} = (100 - 200)$ pF (Murata 81-LXRW19V201-058) such that $C_i = C_F^{(i)} + C_V^{(i)}$. The $C_V^{(i)}$ is controlled by voltage variation, providing a degree of tunability to control the natural frequency of each resonator using an EG&G Instruments 7265 DSP lock-in amplifier that outputs an analog voltage up to a 1 mV resolution (see Fig. 3(b)). Thus, we can precisely control the frequency detuning $\varepsilon$. There are a total of four digital-to-analog (DAC) outputs from the lock-in amplifier and they are each connected to the circuit via a Bayonet Neill–Concelman (BNC) port. Two of these four BNC ports are used to get precise control $C_V^{(i)}$ that detunes the resonant frequency between the $RLC$ tanks. A sub-circuit was built to stabilize our control $C_V^{(i)}$ against unwanted changes in capacitances due to possible voltage fluctuations. It consists of a resistor, $R_V = 4.99$ kΩ, that controls current flow and $C_{BP} = 10$ µF to act as a bypass capacitor. The components used for $R_V$ and $C_{BP}$ are Yageo 603-RC0402FR-074K99L and Murata 81-GCM32EL8EH106KA7L, respectively.

The gain (loss) in the $RLC$ tank $i$ results from an effective resistance $R_i^{-1} = \left(R_L^{(i)}\right)^{-1} + \left(R_{NL}^{(i)}\right)^{-1}$ of the in parallel $R_L^{(i)}$ and $R_{NL}^{(i)}$, where $R_L^{(i)}$ describes the linear behavior of the current-voltage ($I-V$) relationship in a normal resistive element whereas $R_{NL}^{(i)}$ describes the nonlinear behavior due to the presence of back-to-back diodes $D_i$ that is connected in series to $R_{NL}^{(i)}$ (see green and red blocks in Fig. 3(a)). The model used for $R_{NL}^{(i)} = (0-1)$ kΩ is a Bourns 652-3269W-1-102GLF while an Onsemi 512-1N914BWS component is used for $D_i$ to create the nonlinearity present in these $RLC$ tanks. Collectively, $R_{NL}^{(i)}$ and $D_i$ create a nonlinear unit in each $RLC$ tank. To produce gain, an active circuit in $RLC$ tank 1 in the configuration of a negative impedance converter is used. The operational amplifier (op-amp) used to achieve this is model Analog Devices 584-ADA4862-3YRZ-R7. It is powered by a custom DC linear power supply connected to $V_{\pm} = \pm 6$ V. Part of this component houses a voltage divider made up of $R_{G1} = R_{G2} = 550$ Ω that connects from the output of the op-amp to the inverting input of the op-amp. A potentiometer, $R_L^{(1)} = (0-2)$ kΩ, was connected from the output of the op-amp to the non-inverting input of the op-amp to create negative resistance. The model used for $R_L^{(1)}$ is a Bourns 652-3269W-1-202GLF. $R_L^{(1)}$ is connected in-parallel to $R_{NL}^{(1)}$ on one end to collectively produce nonlinear gain. On the other end, it is connected to $C_E$ to capacitively connect it to the transmission line (TL). The component used for $C_E = 10$ pF is a Kemet 80-C0805C100FDTACTU. To produce nonlinear loss, $R_{NL}^{(2)}$, which is identical in components and arrangement to $R_{NL}^{(1)}$, is connected in-parallel on one end to a fixed resistor $R_L^{(2)} = 2.5$ kΩ. On the other end, $R_{NL}^{(2)}$ is connected to another piece of $C_E$ to give the same degree of capacitive coupling to the transmission lines. The transmission line that is capacitively coupled to $R_{NL}^{(2)}$ is connected to the spectrum/vector network analyzer, while the other transmission line connected to $R_{NL}^{(1)}$ is connected to a 50 Ω termination cap.

The sensing platform of these $RLC$ tanks is the capacitive coupling that is sensitive to voltage variations. To achieve this, a pair of in-parallel capacitors, $C_\kappa = \kappa \cdot C \equiv (C_{VC}^{(1)} + C_{VC}^{(2)})$, are used where $C_{VC}^{(i)} = (100 - 200)$ pF (see purple block in Fig. 3(a)). Collectively, the coupling strengths achievable in our experiment is $C_\kappa = (200 - 400)$ pF where $0.18 \leq \kappa \leq 0.36$ which gives us access through $\kappa = \kappa_{NLEPD}$. Each $C_{VC}^{(i)}$ exists within the same sub-circuit structure for $C_V^{(i)}$ with the components of $C_{VC}$ being the same of $C_V^{(i)}$ as well as identical components for $R_V$ and $C_{BP}$. The remaining two out of the four DACs on the lock-in amplifier is used to control the voltages applied to the two $C_{VC}^{(i)}$ units present in the coupling element of our circuit.

## Appendix IV: Temporal Coupled Mode Theory Simulations for Noise Analysis

The TCMT simulations of Eq. (1) were performed using a fourth order Runge-Kutta algorithm for a total evolution time $t_{ev} = 10^5/f_k$, which guarantees the evolution is slow such that the system can adapt to the changes in the parameter space. Meanwhile, $f_k = \nu_k \cdot f_0$, where $f_0$ is the natural frequency of the untuned resonator and $\nu_k$ is the shift to the resonant frequency due to capacitive coupling between the resonators.

Noise was incorporated in two ways: thermal noise in the fields (additive noise) and fluctuations in the values of the parameters $\kappa$ and $\varepsilon$ (multiplicative noise). The standard deviation for the multiplicative noise was set equal to one percent of the instantaneous parameter values. In both cases, the noise sources follow a Gaussian distribution. The routines were implemented using a uniform Park & Miller pseudo-random number generator combined with a shift register. The determination of the bistability cycle width was performed using both the frequency, as in the experimental method, and the relative phase $\varphi$, which also undergoes jumps at the hysteresis loop's edges. For the noise analysis, the Allan deviation was calculated over all the realizations for each one of the $\kappa$ sets in Figs. 5(a,c).